\documentclass[conference]{IEEEtran}
\IEEEoverridecommandlockouts

\def\BibTeX{{\rm B\kern-.05em{\sc i\kern-.025em b}\kern-.08em
    T\kern-.1667em\lower.7ex\hbox{E}\kern-.125emX}}

\usepackage{graphicx}
\usepackage{tikz}
\usepackage{amsmath}

\usepackage{filecontents}
\usepackage{xspace}
\usepackage{url} 
\usepackage[altpo]{backnaur}
\usepackage{booktabs, multirow} %
\usepackage[ruled, vlined, linesnumbered, commentsnumbered, longend]{algorithm2e}

\newcommand{\titlelogo}{\makebox[0pt][r]{\raisebox{-0.20em}{\includegraphics[height=1.18em,trim=70 18 70 18,clip]{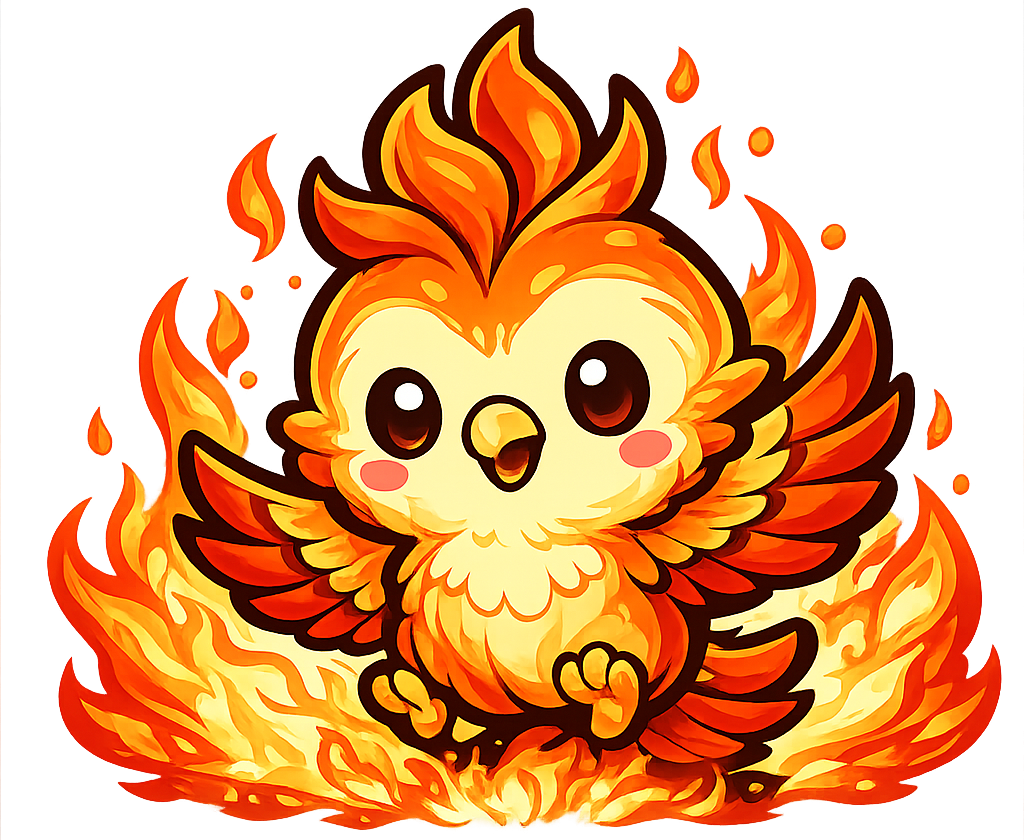}}\hspace{0.16em}}}
\newcommand{\ignore}[1]{}

\definecolor{applegreen}{rgb}{0.55, 0.71, 0.0}

\newcommand{\parabf}[1]{\noindent\textbf{#1}}
\newcommand{\CodeIn}[1]{{\small \texttt{#1}}}

\newcommand{\tactic}{\textsc{Phoenix}\xspace}
\newcommand{\whitefox}{\textsc{WhiteFox}\xspace}
\newcommand{\fuel}{\textsc{FUEL}\xspace}
\newcommand{\titanfuzz}{\textsc{TitanFuzz}\xspace}
\newcommand{\bandit}{Bandit}
\newcommand{\csa}{CSA}

\newcommand{\knighter}{KNighter\xspace}
\newcommand{\fuzzgpt}{\textsc{FuzzGPT}\xspace}

\newcommand{\pt}{PyTorch\xspace}
\newcommand{\jax}{JAX\xspace}
\newcommand{\tvm}{TVM\xspace}

\newcommand{\eg}{\emph{e.g.,}\xspace}
\newcommand{\ie}{\emph{i.e.,}\xspace}

\usepackage{subcaption}
\usepackage{placeins}
\usepackage{xurl}
\usepackage{hyperref}

\begin{document}

\date{}

\title{\texorpdfstring{\titlelogo Rise}{Rise} From The Ashes: LLM-based Static Analysis for Deep Learning Framework Bugs}


\author{
\IEEEauthorblockN{Shaoyu Yang}
\IEEEauthorblockA{\textit{State Key Laboratory for Novel Software Technology} \\
\textit{Nanjing University}\\
Nanjing, China \\
shaoyuyoung@gmail.com}
\and
\IEEEauthorblockN{Chunrong Fang}
\IEEEauthorblockA{\textit{State Key Laboratory for Novel Software Technology} \\
\textit{Nanjing University}\\
Nanjing, China \\
fangchunrong@nju.edu.cn}
\and
\IEEEauthorblockN{Haifeng Lin}
\IEEEauthorblockA{\textit{State Key Laboratory for Novel Software Technology} \\
\textit{Nanjing University}\\
Nanjing, China \\
linhaifeng0716@163.com}
\and
\IEEEauthorblockN{Xiang Chen}
\IEEEauthorblockA{\textit{School of Artificial Intelligence and Computer Science} \\
\textit{Nantong University}\\
Nantong, China \\
xchencs@ntu.edu.cn}
\and
\IEEEauthorblockN{Zhenyu Chen}
\IEEEauthorblockA{\textit{State Key Laboratory for Novel Software Technology} \\
\textit{Nanjing University}\\
Nanjing, China \\
zychen@nju.edu.cn}
}

\author{
\IEEEauthorblockN{
Shaoyu Yang\IEEEauthorrefmark{1},
Haifeng Lin\IEEEauthorrefmark{1},
Chunrong Fang\IEEEauthorrefmark{1}\textsuperscript{\#},
Xiang Chen\IEEEauthorrefmark{2},
Wei Cheng\IEEEauthorrefmark{3},
Jiawei Liu\IEEEauthorrefmark{1},\\
Yiyu Zhang\IEEEauthorrefmark{1},
Hongyu Liu\IEEEauthorrefmark{4},
Zhenyu Chen\IEEEauthorrefmark{1}
}
\IEEEauthorblockA{
\IEEEauthorrefmark{1}State Key Laboratory for Novel Software Technology,
Nanjing University, China
}
\IEEEauthorblockA{
\IEEEauthorrefmark{2}School of Artificial Intelligence and Computer Science, 
Nantong University, China
}
\IEEEauthorblockA{
\IEEEauthorrefmark{3}College of Computer Science and Technology/College of Software, 
Nanjing University of Aeronautics and Astronautics, China
}
\IEEEauthorblockA{
\IEEEauthorrefmark{4}Beijing Academy of Artificial Intelligence, China
}
\IEEEauthorblockA{
shaoyuyang@gmail.com,
linhaifeng0716@163.com,
fangchunrong@nju.edu.cn,\\
xchencs@ntu.edu.cn,
chengweii@nuaa.edu.cn,
jwliu@nju.edu.cn,\\
zhangyy0721@smail.nju.edu.cn,
hyliu@baai.ac.cn,
zychen@nju.edu.cn
}

\IEEEauthorblockA{\textsuperscript{\#}Corresponding author}
}

\maketitle
\thispagestyle{plain}
\pagestyle{plain}

\begin{abstract}
Deep learning (DL) frameworks are critical AI infrastructures that often hide bugs with serious security implications. 
While dynamic approaches such as fuzzing are effective in uncovering these bugs, they require real test execution and incur high computational costs. 
Static analysis is a natural complement because it can detect bugs without runtime execution, offering fast and scalable testing. Unfortunately, there is still limited work targeting static analysis for DL frameworks due to their multilingual architectures and tensor-related program state.

We present \tactic, the first LLM-based static analysis technique for DL frameworks. Our key insight is that cross-language tensor flows in DL frameworks can be modeled, together with concrete code context, as a structured semantic bridge intermediate representation (SBIR) that LLMs can analyze for potential bugs in tensor semantic propagation. 
We implement this insight through a multi-agent workflow. A summarization agent first distills bug summaries from historical bug-fix patches and CWE rules. Guided by each summary, an extraction agent identifies bug-relevant repository symbols for code retrieval, and a generation agent synthesizes grounded SBIRs from the retrieved context. Finally, an analysis agent is leveraged to check SBIRs and report potential bugs.
Our evaluation shows that \tactic is a practical complement to dynamic DL framework testing for bug finding.
To date, \tactic has found 31 real new bugs in \pt for different heterogeneous hardware backends (Intel CPU, NVIDIA CUDA, and Apple MPS). Among them, 20 submitted bug-fixing patches have been merged into upstream.
\end{abstract}


\section{Introduction}
\label{sec:intro}

Deep learning (DL) frameworks, such as \pt~\cite{PyTorch}, \jax~\cite{JAX}, and \tvm~\cite{tvm}, are now critical infrastructure for building, optimizing, and deploying DL and large language model (LLM) applications~\cite{vllm-pagedattention,sglang2024}. Their correctness affects downstream systems in safety-sensitive domains, including autonomous driving~\cite{deeptest}, biometric recognition~\cite{deepface}, and financial services~\cite{finance-survey}. A framework bug can therefore propagate beyond an ordinary library failure because an inconsistent tensor contract, an unsafe backend assumption, or a missing runtime check may silently corrupt numerical results, crash production systems, or expose memory-safety risks~\cite{faulttriggers,chen2025yourcompiler}.

Dynamic testing is a standard software testing paradigm for finding faults through program execution~\cite{ammann2017introduction,jorgensen2013software}. In DL frameworks, fuzz testing (fuzzing) techniques~\cite{cradle,lemon,audee,graphfuzz,nnsmith,neuri} generate or mutate DL programs, execute them against a framework or compiler, and report bugs through crash, differential, numerical, or metamorphic oracles~\cite{nnsmith,whitefox}. Recent fuzzers using LLMs further improve test generation by using LLM knowledge about API syntax and tensor constraints~\cite{titanfuzz,whitefox}. Despite these advances, dynamic techniques still need to construct an executable test that reaches the buggy path and triggers an observable failure. This requirement becomes increasingly costly in modern DL systems, where bug-triggering inputs may depend on large models, heterogeneous backends, or specific Python, C++/CUDA execution paths~\cite{faulttriggers}.

Static analysis~\cite{cousot1977abstract,sadowski2015tricorder} is a natural complement because it can inspect implementation logic before a concrete failure-inducing input is found. However, existing static analyzers~\cite{bandit,clang-static-analyzer} do not directly fit DL frameworks. General tools are usually designed for single programming language or a narrow set of local syntactic rules, while DL framework bugs often arise from inconsistent tensor semantics across Python APIs, C++ dispatchers, and CUDA kernels. Consequently, the community still lacks a practical static analysis technique for detecting DL framework bugs whose root causes lie in cross-language inconsistencies in tensor semantics.

The key challenge is that many DL framework bugs live between language layers rather than inside a single function. As the motivating example in \S~\ref{sec:bg} shows, a frontend API may define a tensor-rank contract, while the backend implementation may consume the same tensor metadata under a different or missing assumption. A Python analyzer can miss the backend consequence, whereas a C/C++ analyzer can miss the frontend contract. Detecting such bugs therefore requires reasoning about how tensor semantics propagate across Python, C++, and CUDA code, which poses two technical challenges:

\textbf{(\textit{C1}) Hybrid multilingual architectures.} High-level Python APIs, C++ dispatchers, CUDA kernels, and backend implementations are connected through foreign function interfaces (FFIs) and framework dispatch mechanisms~\cite{pytorch-neurips2019}. Once control crosses these boundaries, a static analyzer must preserve the source-language contract and the target-language implementation context. Otherwise, the bug-relevant relation disappears and cross-language semantics are lost.

\textbf{(\textit{C2}) Tensor-centric program state.} The relevant program state is tensor-centric rather than scalar-centric. A tensor carries shape, dtype, stride, device, aliasing, layout, allocation, and runtime-state information, and these attributes evolve through framework operators. Bugs often occur when a backend implementation consumes one of these attributes under an assumption that was not established or preserved by earlier layers. Hand-writing rules would require substantial domain-specific expertise and would be difficult to scale across thousands of framework APIs and backend paths~\cite{nnsmith,sadowski2015tricorder}.

These challenges lead to a central question: \textbf{\textit{Can we reason about DL framework bugs at the level of cross-language tensor transfers, rather than at the level of language-specific syntax?}} Our insight is that although DL framework implementations are multilingual and stateful, many bug-relevant contracts reduce to how tensor attributes and guard conditions move from a source entity to a target entity. This transfer-level abstraction helps alleviate the two aforementioned challenges. It bridges language boundaries while retaining tensor-centric state. We therefore introduce a \textit{semantic bridge intermediate representation} (SBIR), which records cross-language semantic transfers together with tensor attributes and repository code context. SBIR abstracts away language-specific syntax while preserving the information needed to reason about shape, dtype, stride, device, aliasing, and guard-state consistency.

Based on SBIR, we present \tactic, an LLM-based static analysis technique for DL frameworks. Rather than asking an LLM to scan the whole codebase, \tactic constrains LLM reasoning around semantic bridges grounded in repository code. This design lets LLMs use their cross-language code understanding (\textbf{addressing \textit{C1}}) where it is most useful~\cite{codescope,swebench}, while keeping the analysis focused on concrete tensor contracts and semantic inconsistencies (\textbf{addressing \textit{C2}}).

In summary, this paper makes the following contributions.

\begin{itemize}
    \item \textbf{Perspective.} We introduce a static analysis perspective for DL framework testing and formulate cross-language tensor-semantic inconsistency as the central bug pattern to be detected without runtime execution.
    \item \textbf{Technique.} We design \tactic, an LLM-based static analysis workflow that combines bug pattern summarization, SBIR generation augmented by retrieval, and semantic analysis over SBIR to detect framework bugs.
    \item \textbf{Evaluation.} We evaluate \tactic on \pt and show that it detects 31 real bugs with substantially fewer false positives than traditional static analyzers. The results also show that \tactic is complementary to LLM-based DL framework fuzzers, finding 30 bugs not detected by the dynamic baselines.
    \item \textbf{Impact.} Among the 31 bugs reported by \tactic, 26 have been confirmed by maintainers, and 20 patches we submitted have been merged into the \pt repository.
\end{itemize}

\section{Background and Motivation}
\label{sec:bg}

\subsection{Tensor Safety in DL Frameworks}

Modern DL frameworks, such as \pt~\cite{PyTorch} and \jax~\cite{JAX}, adopt a hybrid language architecture to balance usability and performance~\cite{pytorch-neurips2019}. A high-level language, typically Python, exposes flexible APIs for model construction, while low-level C++/CUDA backends implement dispatching, memory management, and performance-critical kernels.

\parabf{Tensors as semantic carriers.} The central object crossing these layers is the \textit{tensor}~\cite{pytorch-neurips2019,JAX}. In DL frameworks, a tensor is not merely a multidimensional array. It also carries semantic metadata such as shape, dtype, stride, device, layout, and runtime state. During framework execution, Python APIs prepare or transform this metadata, C++ dispatchers route the operation to backend implementations, and CUDA or device-specific kernels consume the resulting attributes to perform computation. Correct execution, therefore, depends on preserving tensor semantics across language boundaries.

\parabf{Safety risks from broken tensor contracts.} However, this architecture creates a security-relevant semantic gap. The frontend often specifies API-level contracts, such as valid tensor ranks, compatible dtypes, or layout requirements. The backend, optimized for performance, may assume that these contracts have already been checked before it performs pointer arithmetic, memory access, indexing, or device-specific synchronization~\cite{pytorch-neurips2019,faulttriggers}. Bugs arise when a contract established or implied at one layer is not preserved at another layer. In such cases, invalid tensor metadata can reach backend code that is locally \textit{plausible} but globally unsafe. The resulting failures may include out-of-bounds access, invalid memory-state consumption, or crashes under rare backend configurations. 

\subsection{Motivation Example}
\begin{figure}[htbp]
	\centering
 	\includegraphics[width=1\columnwidth]{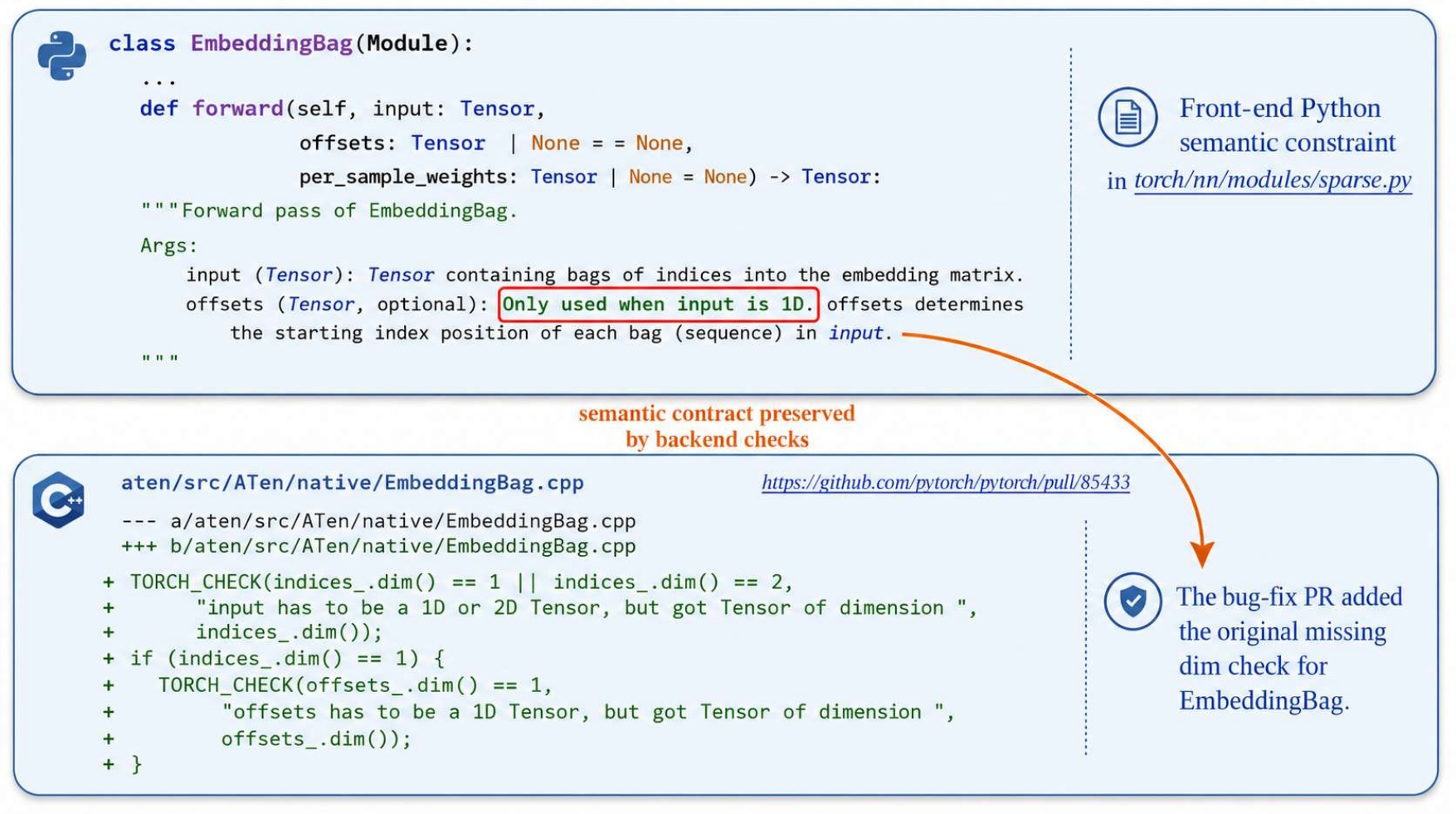}
	\caption{An example of a cross-language bug in \pt.}
	\label{fig:motivation}
\end{figure}

Fig.~\ref{fig:motivation} shows a simplified real-world \pt bug pattern involving \CodeIn{EmbeddingBag}~\cite{pytorch-embeddingbag-issue,pytorch-embeddingbag-doc}. The bug is rooted in a broken tensor-rank contract across the Python and C++ boundary. At the API level, \CodeIn{EmbeddingBag.forward(...)} defines a semantic relation between \CodeIn{input} and \CodeIn{offsets}, where \CodeIn{offsets} is meaningful for one-dimensional \CodeIn{input} only under a corresponding one-dimensional rank constraint. This relation is part of the operator semantics, not a superficial input format preference. In the buggy backend path, the native C++ implementation fails to enforce the same contract in \CodeIn{EmbeddingBag.cpp}. As a result, inconsistent tensor metadata can flow from the frontend into the backend path. Once the backend continues execution under assumptions that only hold for valid 1D/2D inputs and 1D offsets, the operation no longer has a valid semantic basis. A correct backend implementation therefore needs explicit dimension checks that preserve the documented operator contract~\cite{pytorch-embeddingbag-doc}. This case confirms that the failure is \emph{not} a local syntax issue, but a missing preservation of \emph{tensor semantics across language layers}.

\parabf{Why dynamic testing is not sufficient?} Dynamic DL framework testing can expose such bugs when it generates an input that reaches the bug-triggering backend path and triggers an observable failure. However, this depends on constructing the right tensor-rank combination, routing it through the relevant API and dispatch path, and observing a failure under a specific runtime configuration. Many tensor-semantic bugs are latent as they encode inconsistent assumptions that may not crash immediately or may only surface under unusual backend states~\cite{faulttriggers}. This motivates a complementary static view that can inspect semantic preservation without first constructing a concrete failure-inducing test.

\parabf{Why existing static analyzers miss it?} Bandit~\cite{bandit} and CSA~\cite{clang-static-analyzer} fail on this case for different reasons. Bandit is limited to the Python frontend. It can inspect the API-level code, but it cannot verify whether the rank contract is preserved after execution enters the native backend. CSA can analyze the C++ implementation, but it does not know the frontend semantic contract that \CodeIn{input} should be 1D/2D and that \CodeIn{offsets} should be 1D in the 1D-input case. Thus, Bandit loses the backend consequence, while CSA loses the frontend semantics. The bug lies in the transfer between these views.

\parabf{Design challenges.} The example suggests three requirements for DL framework static analysis. First, the analysis should represent tensor metadata, not only variable names or scalar values. Second, it should connect source and target entities across Python, C++, CUDA, and backend-specific code. Third, it should stay grounded in repository code context, because unconstrained reasoning over a large codebase can easily invent plausible but unsupported code paths. Such requirements motivate the design of \tactic in \S~\ref{sec:appro}.

\section{\tactic Approach}
\label{sec:appro}

\begin{figure*}[htbp]
	\centering
	\includegraphics[width=0.8\textwidth]{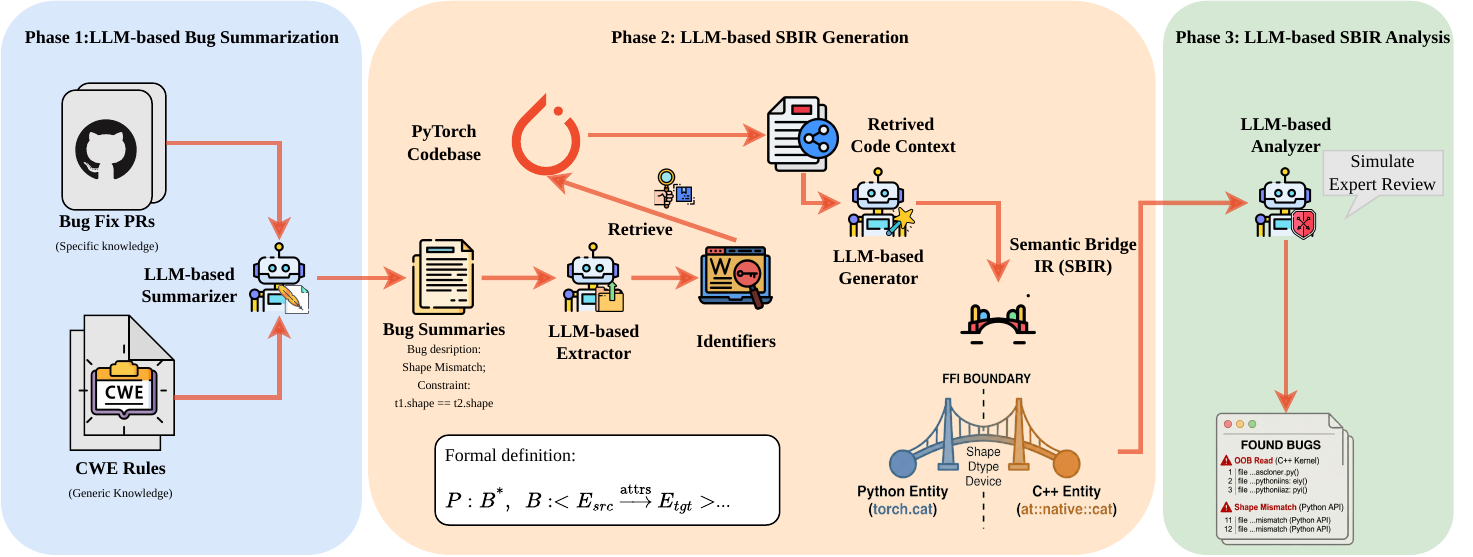}
	\caption{Overview of \tactic}
	\label{fig:overview}
\end{figure*}

\textbf{Overview.} Fig.~\ref{fig:overview} shows the workflow of \tactic, which uses a curated bug dataset as external knowledge and coordinates four LLM agents across three phases. The workflow is centered on SBIR, a domain-specific intermediate representation that captures the cross-language propagation of tensor semantics in DL frameworks. In the first phase, a \textbf{summary agent} normalizes bug knowledge into reusable bug summaries. In the second phase, an \textbf{extraction agent} and a \textbf{generation agent} retrieve code context from the repository and synthesize SBIR. In the third phase, an \textbf{analysis agent} reasons over SBIR to report candidate bugs. The preceding dataset construction step prepares the knowledge source used by these phases. We introduce each component in the following subsections.

\subsection{Bug Dataset Collection}
\label{subsec:bug-collection}
\tactic requires bug knowledge to guide both SBIR construction and semantic analysis. However, existing public datasets are mostly designed for general-purpose software or single-language programs~\cite{sard-juliet,bigvul,devign}. They rarely capture bugs specific to DL frameworks. We therefore build a dedicated bug dataset as the knowledge source of \tactic.

\begin{table}[t]
\centering
\caption{Bug dataset used by \tactic for bug-pattern summarization.}
\label{tab:bugdataset}
\footnotesize
\setlength{\tabcolsep}{3pt}
\begin{tabular}{@{}p{.18\columnwidth}p{.62\columnwidth}r@{}}
\toprule
\textbf{Source} & \textbf{Bug or weakness type} & \textbf{\# Items} \\
\midrule
\multirow{4}{*}{Bug-fix PRs}
& Tensor metadata and layout contract violations & 10 \\
& Predicate and API-contract validation errors & 10 \\
& Resource, pointer, and runtime-state errors & 10 \\
& Heterogeneous backend and concurrency errors & 10 \\
\midrule
\multirow{16}{*}{CWE rules}
& Missing Authorization (CWE-862) & 1 \\
& Out-of-bounds Write (CWE-787) & 1 \\
& Path Traversal (CWE-22) & 1 \\
& Use After Free (CWE-416) & 1 \\
& Out-of-bounds Read (CWE-125) & 1 \\
& OS Command Injection (CWE-78) & 1 \\
& Code Injection (CWE-94) & 1 \\
& Classic Buffer Overflow (CWE-120) & 1 \\
& Dangerous file upload (CWE-434) & 1 \\
& NULL Pointer Dereference (CWE-476) & 1 \\
& Deserialization of Untrusted Data (CWE-502) & 1 \\
& Heap-based Buffer Overflow (CWE-122) & 1 \\
& Incorrect Authorization (CWE-863) & 1 \\
& Improper Input Validation (CWE-20) & 1 \\
& Improper Access Control (CWE-284) & 1 \\
& Unbounded resource allocation (CWE-770) & 1 \\
\midrule
\multicolumn{2}{@{}r}{Total} & 56 \\
\bottomrule
\end{tabular}
\end{table}

We collect the dataset from two complementary sources. The first source is historical bug-fix PRs from the \pt repository~\cite{pytorch-tensor-prs}. The first and second authors searched \pt PRs with the keyword \CodeIn{tensor}, screened the merged PRs for tensor-related bug fixes, and removed changes that do not represent bug fixes, including documentation updates, formatting changes, pure refactoring, and pure feature modifications. The retained PRs were grouped into four common tensor bug types that repeatedly appear in DL framework implementations. These types are tensor metadata and layout contract violations, predicate and API contract validation errors, resource, pointer, and runtime state contract errors, and heterogeneous backend or concurrency state errors. For each type, we ordered candidate PRs by time and selected ten PRs following this chronological order. This process yields 40 bug-fix PRs, which is sufficient for generating enough SBIRs.

The second source is the \textit{2025 CWE Top 25 Most Dangerous Software Weaknesses}~\cite{cwe-top-25-2025}. CWE entries provide general security knowledge that may not have appeared in our collected \pt patches but can still guide the search for security-relevant risks. We retain 16 CWE rules that can be mapped to DL framework implementation risks and discard weakness categories that are difficult to relate to tensor semantics, such as database-specific injection issues. Table~\ref{tab:bugdataset} summarizes the final dataset. These items are not treated as detection results. Instead, they are used as source knowledge for the summarizer, which uses an LLM to convert heterogeneous PR and CWE descriptions into normalized bug patterns.



\newcommand{\decNT}[1]{{\sf #1}\xspace}
\newcommand{\expr}{expr\xspace}
\newcommand{\rinp}{\mathcal I \xspace}
\newcommand{\dinp}{I \xspace}
\newcommand{\attr}{A\xspace}
\subsection{Semantic Bridge IR}
\label{subsec:sbir}

\textbf{Design scope and extensibility.} SBIR is designed as a lightweight semantic abstraction rather than a complete program IR\footnote{We use the term ``IR'' in the compiler sense because this semantic carrier is an internal artifact in the workflow, rather than a user-visible input or output.}. Its goal is not to encode every execution detail of \pt, but to preserve tensor-level contracts that are relevant to cross-language bugs. \tactic therefore focuses on recurring semantic transfer types, including data propagation, aliasing, dispatch, guard checking, control dependency, metadata conversion, mutation, and allocation. These types are sufficient to express the bug patterns studied in this work, including missing dimension checks, dtype/device inconsistencies, unsafe layout assumptions, and backend precondition violations. SBIR is intentionally extensible, so new transfer types and metadata keys can be added when \tactic is applied to new framework components or new bug classes.

\textbf{Formal definition.} Formally, a \pt program execution is represented as a sequence of semantic bridges that model how a source-language tensor entity is mapped to a target-language entity together with its transferred semantic attributes and code context. The formal syntax of the semantic bridge IR is given below.

{\small
\begin{bnf*}
  \bnfprod{\decNT{P}}
    {\bnfsp \bnfpn{\decNT{$B*$}}}\\
  \bnfprod{\decNT{B}}
    {\bnfts{$\langle$}\bnfsp
     \bnfpn{\decNT{$E_{src}$}}\bnfts{,}\bnfsp
     \bnfpn{\decNT{$E_{dst}$}}\bnfts{,}\bnfsp
     \bnfpn{\decNT{T}}\bnfts{,}\bnfsp
     \bnfpn{\decNT{$\Phi$}}\bnfts{,}\bnfsp
     \bnfpn{\decNT{$\Gamma$}}\bnfsp
     \bnfts{$\rangle$}}\\[1ex]
  \bnfprod{\decNT{E}}
    {\bnfpn{\decNT{$\ell$}}\bnfts{.}\bnfpn{\decNT{e}}%
     \bnfts{$[$}\bnfsp\bnfts{$.$}\bnfpn{\decNT{f}}\bnfsp\bnfts{$]$}}\\
  \bnfprod{\decNT{$\ell$}}
    {\bnfts{Python}\bnfor
     \bnfts{CPP}\bnfor
     \bnfts{CUDA}
    }\\
   \bnfprod{\decNT{T}}
   {\bnfts{data}\bnfor
    \bnfts{alias}\bnfor
    \bnfts{grad}} \\
    \bnfmore{\bnfts{dispatch}\bnfor
    \bnfts{guard}\bnfor
    \bnfts{control}} \\
    \bnfmore{
    \bnfts{metadata}\bnfor
    \bnfts{mutation}\bnfor
    \bnfts{allocation}
   }\\
   \bnfprod{\decNT{$\Phi$}}
   {\bnfsp \bnfpn{\decNT{$\phi*$}}}\\
   \bnfprod{\decNT{$\phi$}}
   {\bnfpn{\decNT{k}}\bnfsp\bnfts{=}\bnfsp\bnfpn{\decNT{v}}}\\
   \bnfprod{\decNT{k}}
   {\bnfts{dtype}\bnfor
    \bnfts{shape}\bnfor
    \bnfts{stride}} \\
    \bnfmore{\bnfts{layout}\bnfor
    \bnfts{device}\bnfor
    \bnfts{req\_grad}} \\
    \bnfmore{\bnfts{alias}\bnfor
    \bnfts{capacity}\bnfor
    \bnfts{state}
   }\\
   \bnfprod{\decNT{e}}
   {\bnfts{id}}\\
   \bnfprod{\decNT{f}}
   {\bnfts{id}}\\
   \bnfprod{\decNT{v}}
   {\bnfts{id}\bnfor
    \bnfts{num}\bnfor
    \bnfts{bool}\bnfor
    \bnfts{[}\bnfts{num}\bnfts{$^{*}$}\bnfts{]}}\\
    \bnfmore{\bnfts{expr}\bnfor
    \bnfts{pred}}\\
    \bnfprod{\decNT{$\Gamma$}}
   {\bnfsp \bnfpn{\decNT{$\gamma*$}}}\\
   \bnfprod{\decNT{$\gamma$}}
    {\bnfts{filename}\bnfsp\bnfts{=}\bnfsp\bnfts{code}}\\
\end{bnf*}
}

A program $\decNT{P}$ is a sequence of semantic bridges $\decNT{B}$. Each bridge $\decNT{B}$ is a five-tuple $\langle\decNT{E_{src}}, \decNT{E_{dst}}, \decNT{T}, \decNT{\Phi}, \decNT{\Gamma}\rangle$ that describes one semantic transfer from a source entity to a destination entity. The fields $\decNT{E_{src}}$ and $\decNT{E_{dst}}$ are instances of the entity nonterminal $\decNT{E}$. An entity $\decNT{E}$ has the form $\decNT{\ell}.\decNT{e}[.\decNT{f}]$, where $\decNT{\ell}$ identifies the language layer, $\decNT{e}$ denotes a program object such as an API, function argument, tensor, pointer, kernel parameter, or class instance, and the optional $\decNT{f}$ denotes the specific member, property, or metadata item of interest, such as shape, dtype, device, size, data pointer, index, or bounds. If the analysis concerns the object as a whole, $\decNT{f}$ is omitted. The language-layer nonterminal $\decNT{\ell}$ ranges over the terminals \CodeIn{Python}, \CodeIn{CPP}, and \CodeIn{CUDA}. The nonterminal $\decNT{T}$ records the transfer type, such as data propagation, alias propagation, gradient forwarding, dispatch, guard checking, control dependency, metadata conversion, mutation, or allocation. $\decNT{\Phi}$ is a set of semantic constraints. Each constraint $\decNT{\phi}$ has the form $\decNT{k} = \decNT{v}$, where $\decNT{k}$ is an attribute key and $\decNT{v}$ is an attribute value. The value $\decNT{v}$ can be a literal, identifier, numeric vector, symbolic expression, or predicate, which lets SBIR represent both tensor metadata and code-derived constraints such as capacity expressions and guard conditions. These keys record tensor properties and runtime states that should be preserved or intentionally changed during propagation, including shape, dtype, stride, layout, device, aliasing status, capacity, and state. $\decNT{\Gamma}$ is a set of code-context entries. Each entry $\decNT{\gamma}$ has the terminal form \CodeIn{filename = code} and stores the repository code context of the bridge. This context lets the analyzer interpret the bridge with its surrounding \pt implementation.

\textbf{Advantages of SBIR Design.} SBIR records only bug-relevant semantic transfers rather than full program behavior. This choice keeps the abstraction compact enough for reasoning with LLMs while retaining the tensor attributes, guard conditions, and repository context needed for bug analysis. SBIR gives \tactic a common analysis surface across the language layers modeled in this work, namely Python, C++, and CUDA. Instead of asking the analyzer to compare heterogeneous code fragments directly, \tactic checks whether the semantic contracts encoded in SBIR are preserved as tensor information moves across these language layers.

\subsection{LLM-based Bug Summarization}
\label{subsec:summary}
The raw bug dataset described in \S~\ref{subsec:bug-collection} is useful as external knowledge, but it is not directly suitable for SBIR construction. Bug-fix PRs contain concrete patches and development context, but they are often verbose and tied to one specific implementation~\cite{tufano2018bugfix}. CWE rules provide general weakness knowledge, but they are too abstract to directly identify tensor-level contracts in DL frameworks. Feeding these raw sources into later stages would make retrieval unstable and would leave the generator with ambiguous bug intent.

\begin{figure}[t]
    \centering
    \includegraphics[width=.86\columnwidth]{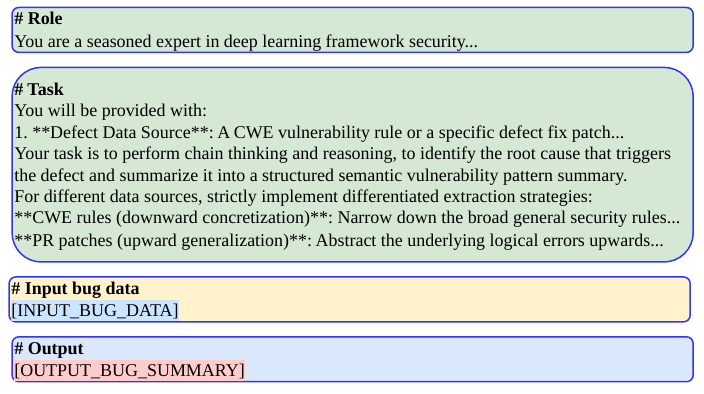}
    \caption{Prompt template for bug-pattern summarization.}
    \label{fig:prompt_summary}
\end{figure}

\tactic therefore first normalizes each source item into a structured bug summary. Fig.~\ref{fig:prompt_summary} shows the summarization prompt. The prompt assigns the LLM the role of a DL framework security expert and asks it to summarize the root cause that triggers the bug. Importantly, the prompt uses different abstraction directions for different data sources. For CWE rules, LLM performs downward concretization by narrowing a general weakness into a semantic pattern for DL frameworks. For PR patches, LLM performs upward generalization by abstracting the concrete patch into a reusable bug pattern. The output \CodeIn{[OUTPUT\_BUG\_SUMMARY]} records the bug type, trigger condition, violated semantic constraint, and relevant tensor or code entities. This summary becomes the semantic anchor for both identifier extraction and SBIR generation.

\subsection{LLM-based SBIR Generation}
\label{subsec:generation}
After bug summarization, \tactic needs to construct an SBIR that connects the bug pattern to actual code in the target repository. A direct strategy would ask an LLM to locate relevant code and generate SBIR in one pass. This strategy is unreliable because the LLM must simultaneously search a large multilingual codebase and synthesize a formal abstraction. In our preliminary investigation, this often led to hallucinated code contexts~\cite{zhang2025siren}, such as non-existent file paths, unsupported tensor attributes, or bridges that were not grounded in the repository.

To avoid this failure mode, \tactic separates code grounding from SBIR generation through an Extract-Retrieve-Generate (ERG) strategy. Algorithm~\ref{algo:search-extraction} gives the procedure. Given raw bug data, a normalized bug summary, and the target codebase, ERG first retrieves repository code contexts and then allows SBIR generation only when such contexts are available.

\begin{figure}[t]
    \centering
    \includegraphics[width=.86\columnwidth]{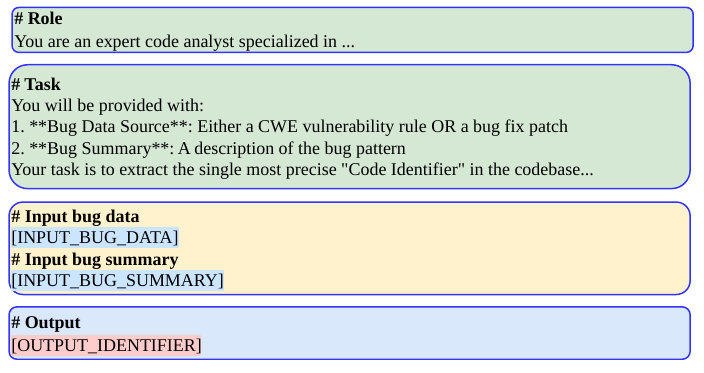}
    \caption{Prompt template for identifier extraction.}
    \label{fig:prompt_extract}
\end{figure}

\begin{algorithm}[t]
\small
\caption{ERG Strategy for SBIR Generation}
\label{algo:search-extraction}
\DontPrintSemicolon
\SetKwProg{Fn}{Function}{:}{}

\SetKwData{bugdata}{bugData}
\SetKwData{bugsumm}{bugSummary}
\SetKwData{identifier}{identifier}
\SetKwData{codebase}{codebase}
\SetKwData{contextset}{contextSet}
\SetKwData{attempts}{attempts}
\SetKwData{sbir}{SBIR}

\SetKwFunction{irgen}{IRGen}
\SetKwFunction{llmg}{Gen}
\SetKwFunction{llme}{Extract}
\SetKwFunction{retrieve}{CodebaseRetrieve}

\Fn{\irgen{\bugdata,\bugsumm,\codebase}}{
    $\attempts, \contextset \leftarrow 0, \emptyset$ \\ \label{algo:init}
    \While{$\attempts < 3$ $AND$ $\contextset = \emptyset$}{ \label{algo:attemptbudget}
        \textcolor{blue}{\tcp{Identifier extraction with LLMs}}
        $\identifier \leftarrow \llme{\bugdata,\bugsumm}$ \\ \label{algo:identifier}
        $\contextset \leftarrow \retrieve{\identifier,\codebase}$ \\ \label{algo:retrieve}
        $\attempts \leftarrow \attempts + 1$ \label{algo:increment}
    }
    \If{$\contextset = \emptyset$}{ \label{algo:contextcondition-start}
        \Return $Null$ \label{algo:contextcondition-end}
    }
    \textcolor{blue}{\tcp{SBIR generation with LLMs}}
    $\sbir \leftarrow \llmg{\bugdata,\bugsumm,\contextset}$ \\ \label{algo:llmgen}
    \Return \sbir \label{algo:return-sbir}
}
\end{algorithm}

Algorithm~\ref{algo:search-extraction} starts by initializing the retry counter and the retrieved context set (Line~\ref{algo:init}). While no code context has been found and the retry budget has not been exhausted (Line~\ref{algo:attemptbudget}), the extractor predicts one identifier from the raw bug data and its summary (Line~\ref{algo:identifier}). The extractor prompt in Fig.~\ref{fig:prompt_extract} asks the LLM to output the single most precise \CodeIn{[OUTPUT\_IDENTIFIER]} for repository search. This identifier can be a Python API, C++ function, CUDA kernel, helper routine, or other concrete symbol in the target codebase. For a bug item like the \CodeIn{EmbeddingBag} case in \S~\ref{sec:bg}, the extractor converts the summarized bug intent into an operator- or backend-related search symbol, which confines retrieval to relevant code contexts instead of reasoning over the whole repository. \tactic then passes the identifier to \CodeIn{CodebaseRetrieve} to obtain function-level code contexts (Line~\ref{algo:retrieve}). If retrieval fails, the loop increments the retry counter (Line~\ref{algo:increment}) and tries another identifier. This retry logic gives the extractor a bounded chance to recover from an unresolved or overly broad identifier. If no context is retrieved after the budget is exhausted, \tactic returns \CodeIn{Null} (Lines~\ref{algo:contextcondition-start}--\ref{algo:contextcondition-end}). This failure branch is important because it prevents the generator from synthesizing SBIR without retrieved repository code context. Only when \CodeIn{contextSet} is non-empty does the generator receive the bug data, bug summary, and retrieved context to generate SBIR (Line~\ref{algo:llmgen}), which is then returned to the downstream analyzer (Line~\ref{algo:return-sbir}).

\begin{figure}[t]
    \centering
    \includegraphics[width=.86\columnwidth]{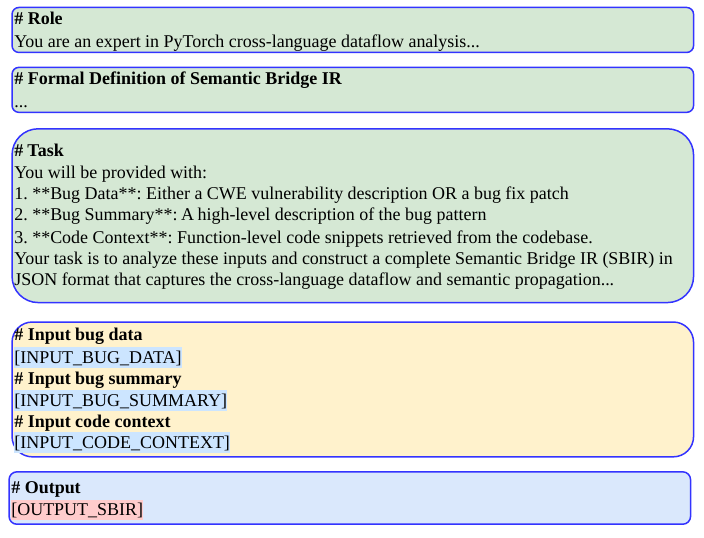}
    \caption{Prompt template for SBIR generation.}
    \label{fig:prompt_generation}
\end{figure}

Fig.~\ref{fig:prompt_generation} shows the generation prompt, where the generator receives the formal definition of SBIR, the original bug data, the normalized bug summary, and the retrieved \CodeIn{[INPUT\_CODE\_CONTEXT]}. It then outputs \CodeIn{[OUTPUT\_SBIR]} in JSON format to describe bug-relevant semantic bridges. This design is intentionally constrained because the generator does not search the codebase by itself and only abstracts from retrieved contexts at function granularity, which keeps the generated SBIR grounded in the actual \pt codebase.

\subsection{LLM-based SBIR Analysis}
\label{subsec:analysis}
The final stage decides whether a generated SBIR indicates a real bug. Although SBIR is structured, analyzing it with manual rules is still difficult. DL framework bugs often depend on how several constraints interact across language boundaries, such as shape checks in Python, layout assumptions in C++, and runtime states in CUDA code. Encoding all such combinations as deterministic rules would require substantial domain effort and would still miss context-dependent logic errors.

\begin{figure}[t]
    \centering
    \includegraphics[width=.86\columnwidth]{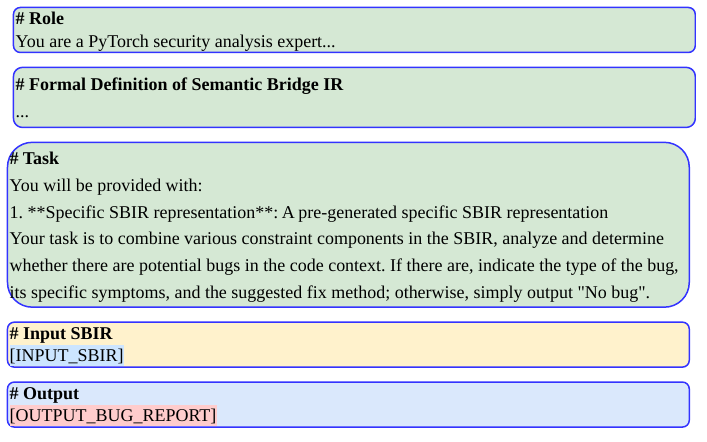}
    \caption{Prompt template for bug analysis over SBIR.}
    \label{fig:prompt_analysis}
\end{figure}

\tactic uses an LLM-based analyzer over SBIR as a semantic trace. As shown in Fig.~\ref{fig:prompt_analysis}, the prompt provides the formal SBIR definition and the concrete \CodeIn{[INPUT\_SBIR]}. The analyzer examines whether the semantic constraints in $\decNT{\Phi}$ are consistently preserved from $\langle\decNT{E_{src}}\rangle$ to $\langle\decNT{E_{dst}}\rangle$ under the code context $\langle\decNT{\Gamma}\rangle$. If a violation is found, it outputs \CodeIn{[OUTPUT\_BUG\_REPORT]} with the bug type, concrete symptom, and suggested fix. If the SBIR is consistent, the analyzer is instructed to output \CodeIn{No bug}. This conservative output keeps the final report focused on semantic inconsistencies with high confidence rather than unconstrained LLM speculation.

\FloatBarrier

\section{Evaluation}
\label{sec:eval}

\subsection{Experiment Setup}
\label{sec:setup}
We design the following research questions (RQs) to evaluate the effectiveness of \tactic:

\parabf{RQ1.} Can \tactic outperform static approaches?  

\parabf{RQ2.} Is \tactic orthogonal to dynamic approaches?

\parabf{RQ3.} What kinds of bugs can \tactic detect?

\parabf{RQ4.} How does each component contribute to \tactic?  


\parabf{Systems under test.} e use \pt~\cite{PyTorch} as the target system because it is a widely used DL framework with a hybrid Python/C++/CUDA architecture, heterogeneous backends, and an active bug-fix history. These properties make it a representative subject for studying cross-language tensor-semantic bugs, and we evaluate \tactic on its nightly version.

\parabf{Metrics.} As we evaluate our approach on the \textit{real-world} \pt codebase, we use two evaluation metrics. \textit{(i) False positive} measures the proportion of analysis reports that are not actual bugs, reflecting the precision of each method. \textit{(ii) Bug detection} counts the number of \textit{new} \pt bugs detected by \tactic and baselines.

\parabf{Baselines.} We organize the baselines into static techniques for RQ1 and dynamic techniques for RQ2.

\textit{Static techniques.} Since there is no existing static analyzer designed specifically for DL framework tensor-semantic bugs, we compare \tactic with two traditional static analysis tools that cover different language layers. \textbf{Bandit}~\cite{bandit} is a Python-oriented static analysis tool that inspects source code for common security issues, making it suitable for the Python frontend components of \pt where high-level API logic is implemented. \textbf{Clang Static Analyzer (CSA)}~\cite{clang-static-analyzer} is a path-sensitive static analysis framework for C/C++ and also supports CUDA, making it suitable for native operators, backend kernels, and low-level control-flow or memory checks. We run both tools as full scans on the language they support.

\textit{Dynamic techniques.} To study orthogonality between static reasoning and dynamic testing, we compare \tactic with two DL fuzzers that use LLMs. \textbf{\titanfuzz}~\cite{titanfuzz} uses LLMs to generate DL programs by constructing seed programs and filling API parameters with context-aware values. \textbf{\whitefox}~\cite{whitefox} is a white-box fuzzer that inspects compiler source code, extracts optimization patterns with an analysis LLM, and uses these patterns to guide test generation. For a fair comparison, we follow the default configurations in their respective papers and replace their base LLM with the same default LLM used by \tactic.

\parabf{Ablation variants.} We design four ablation variants of \tactic for our ablation studies.
\begin{itemize}
    \item \textbf{w/o CWE.} This variant excludes the general weakness patterns derived from CWE. By relying solely on historical PRs, we assess the necessity of high-level, generic weakness patterns in guiding the discovery of bug types.
    \item \textbf{w/o PR.} This variant discards the bug fix PRs. It relies exclusively on CWE rules, allowing us to evaluate the importance of leveraging evolution and historical regression patterns specific to DL frameworks.
    \item \textbf{w/o summary.} This variant omits the bug summarization step. Consequently, the subsequent SBIR generation and keyword search phases operate without the semantic guidance provided by natural language summaries, testing the value of abstracting code changes into semantic intents.
    \item \textbf{w/o ERG.} This variant removes retrieved repository code contexts and directly asks the LLM generator to synthesize SBIRs from the bug data and summary. It evaluates whether generating SBIR without real code context causes hallucinated contexts and substantially more false positives.
\end{itemize}

\parabf{LLM and repetition settings.} We use GPT-5.4 as the default LLM backend. To reduce the effect of LLM nondeterminism during SBIR construction, each PR or CWE item is processed 10 times, producing 10 SBIRs for subsequent analysis.

\parabf{Environment configuration.} Our experiments were conducted on Ubuntu 20.04 LTS with 2 NVIDIA V100 GPUs.

\subsection{RQ1. Superiority over static methods} 
\begin{table}[t]
\centering
\caption{Comparison with traditional static analysis tools.}
\label{tab:rq1}
\small
\begin{tabular}{@{}lrrr@{}}
\toprule
\textbf{Technique} & \textbf{\# Bugs} & \textbf{\# False Alarms (\%)} & \textbf{\# Alarms} \\
\midrule
\tactic & \textbf{31} & \textbf{5 (13.89\%)} & \textbf{36} \\
\bandit & 1 & 171 (99.41\%) & 172 \\
\csa & 0 & 218 (100.00\%) & 218 \\
\bottomrule
\end{tabular}
\end{table}


We compare \tactic against two representative traditional static analysis tools, Bandit~\cite{bandit} for Python and CSA~\cite{clang-static-analyzer} for C/C++ and CUDA. Table~\ref{tab:rq1} summarizes the results. \tactic reports 36 alarms, among which 31 are real \pt bugs. This corresponds to only 5 false alarms and a false positive rate of 13.89\%. In contrast, Bandit reports 172 alarms but only one of them is a real bug, leading to 171 false alarms and a false positive rate of 99.41\%. CSA reports 218 alarms, all of which are false positives after manual validation.

\parabf{Alarm precision.} The results show that traditional static tools produce many alarms but provide little useful signal for DL framework bug detection. Bandit is designed for Python security auditing, so it frequently flags syntactic patterns that are harmless in the \pt execution context, such as framework-internal command construction or configuration handling. CSA reasons over low-level C/C++ and CUDA code, but it does not understand the tensor-level contracts established by the Python frontend and the dispatcher. As a result, it flags many local memory or control-flow risks that are either unreachable under framework invariants or unrelated to tensor semantic bugs. Such high false positive rates make it impractical to use these traditional static tools for DL framework analysis. In contrast, \tactic's design based on SBIR allows it to focus on cross-layer semantic constraints that are more directly related to real bugs, leading to a much lower false positive rate.

\parabf{Bug detection.} Beyond precision, \tactic also detects substantially more real bugs than the traditional tools. Bandit finds only one real bug, and CSA finds none. In comparison, \tactic detects 31 real bugs. This gap comes from the fact that many \pt bugs are not isolated single-language rule violations. They arise when tensor semantics (\eg shape and dtype) are transferred across Python, C++, and CUDA code. By representing these transfers with SBIR, \tactic can check whether backend implementations preserve the semantic constraints expected by higher-level framework logic. This makes \tactic better suited to DL framework analysis than directly applying general-purpose static analyzers.

\parabf{Answer to RQ1.} \tactic substantially outperforms traditional static analyzers on \pt. It detects 31 real bugs with a 13.89\% false positive rate, whereas Bandit and CSA find at most one real bug and produce nearly all false alarms.

\subsection{RQ2. Orthogonality to dynamic methods}
To investigate whether \tactic merely reproduces the findings of existing dynamic testing tools or contributes a complementary perspective, we conduct an orthogonality analysis against state-of-the-art DL fuzzers that use LLMs (\titanfuzz~\cite{titanfuzz} and \whitefox~\cite{whitefox}). The goal of this RQ is not to show that static analysis should replace dynamic testing. Instead, we study whether static reasoning over SBIR and dynamic fuzzing expose different parts of the DL framework bug space. We analyze the results from two perspectives.

\subsubsection{Bug overlap analysis}
\begin{figure}[t]
    \centering
    \includegraphics[width=.95\columnwidth]{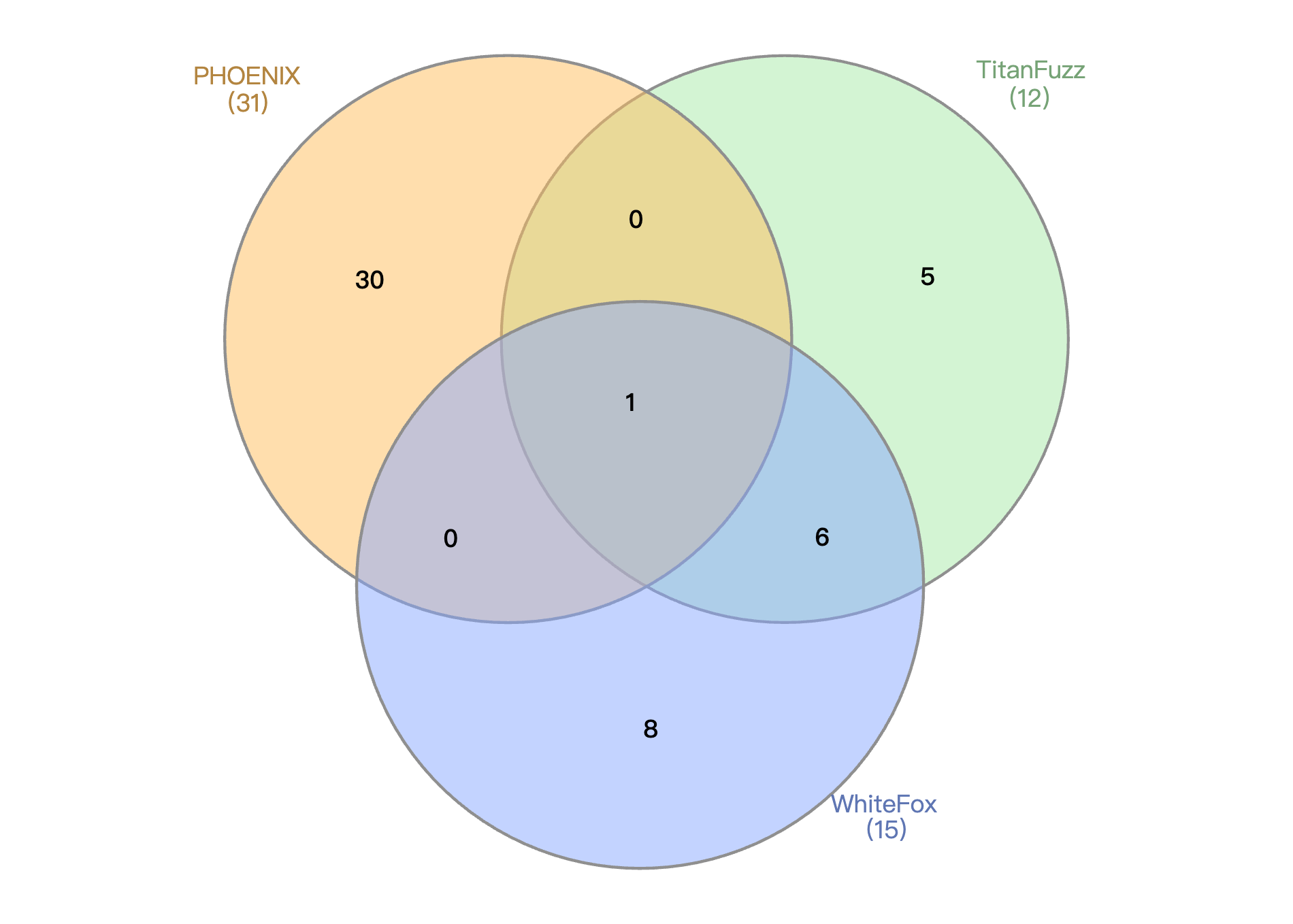}
    \caption{Bug overlap among \tactic and dynamic baselines.}
    \label{fig:rq2-bug-overlap}
\end{figure}

Fig.~\ref{fig:rq2-bug-overlap} shows a small overlap between \tactic and the dynamic baselines. \tactic detects 31 bugs, while \titanfuzz and \whitefox detect 12 and 15 bugs, respectively. The two dynamic baselines together cover 20 distinct bugs, among which 7 are shared by both dynamic methods. In contrast, only one bug overlaps between \tactic and either dynamic baseline, and this bug is found by all three methods. As a result, 30 of the 31 bugs detected by \tactic are unique, while 19 bugs detected by the dynamic baselines are not detected by \tactic. This result directly supports the orthogonality between the two testing paradigms. Dynamic fuzzers are effective when they can generate an input that reaches the buggy path and exposes an observable failure. \tactic instead reasons over source-level semantic transfers encoded in SBIR, allowing it to flag inconsistencies without constructing executable tests. The two designs therefore favor different regions of the bug space, rather than one simply subsuming the other.

\subsubsection{Bug type distribution analysis}
\begin{figure}[t]
    \centering
    \includegraphics[width=\columnwidth]{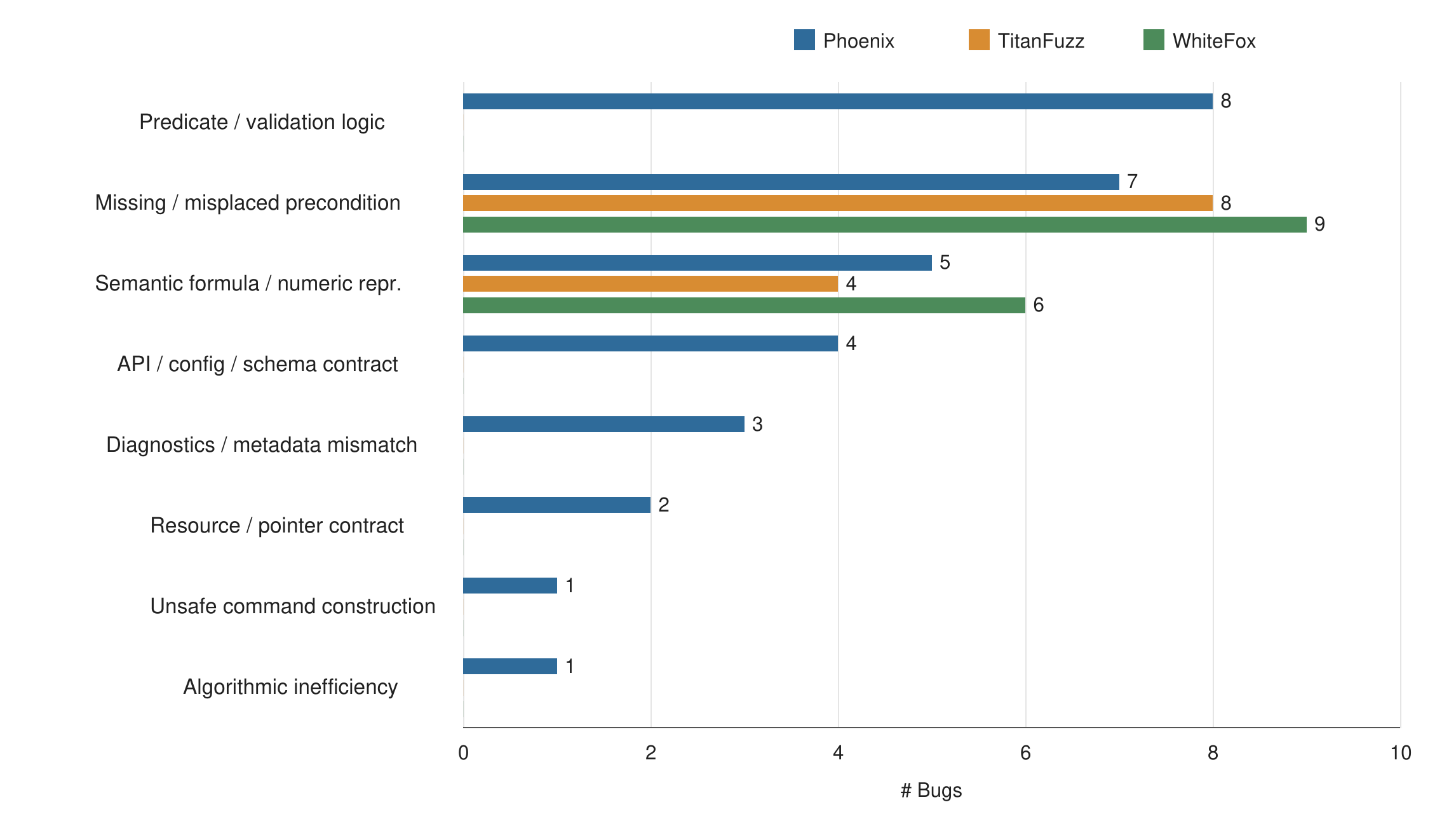}
    \caption{Root-cause distribution of bugs detected by \tactic and dynamic baselines.}
    \label{fig:rq2-root-cause-distribution}
\end{figure}

We further analyze the detected bugs based on their root causes, as illustrated in Fig.~\ref{fig:rq2-root-cause-distribution}. The dynamic baselines are concentrated in two categories (\ie \textit{missing or misplaced preconditions}, and \textit{semantic formula or numeric representation errors}). This concentration is expected because dynamic testing relies on runtime oracles~\cite{barr2014oracle,chen2018metamorphic}. It is naturally strong at exposing bugs that lead to crashes, exceptions, or incorrect numerical results. In contrast, \tactic covers a wider range of root causes, including predicate and validation logic, API and schema contracts, diagnostic metadata, resource or pointer contracts, unsafe command construction, and algorithmic inefficiency. These bugs often involve inconsistent semantic assumptions across framework layers, and their symptoms may depend on rare configurations, backend-specific states, or contracts that are difficult to express as dynamic oracles. The distribution therefore explains the limited overlap in Fig.~\ref{fig:rq2-bug-overlap}. Dynamic methods provide concrete failure-inducing tests, while \tactic contributes a static view that identifies latent semantic inconsistencies.

\parabf{Answer to RQ2.} \tactic is complementary to existing dynamic DL framework testing methods. It contributes 30 bugs not found by the dynamic baselines, while the dynamic baselines contribute 19 bugs not found by \tactic. Together, the three methods cover 50 distinct bugs in our study. This confirms that static analysis over SBIR and dynamic fuzzing should be viewed as mutually reinforcing techniques for DL framework bug detection.

\subsection{RQ3. Bug study}
\pt supports multiple backends and framework components, where tensor semantics are consumed by very different implementations such as CPU runtime code, CUDA kernels, and MPS backend code. Overall, \tactic has produced 31 bug reports for \pt. Among them, 26 have been confirmed by maintainers. After detecting these bugs, we submitted patches to the upstream \pt repository to fix them directly, and 20 patches have been merged. Table~\ref{tab:bug-study} summarizes the reported, confirmed, and fixed bugs across backends and components. 
The results show that the detected bugs are not concentrated in one implementation layer. This distribution supports our claim that SBIR provides an abstraction for analyzing cross-layer tensor semantics without depending on a specific backend.
Because \tactic does not execute \pt workloads during detection, it can inspect heterogeneous backend code without requiring access to every corresponding hardware device.

\begin{table}[t]
\centering
\caption{Bug reports, confirmed bugs, and merged fixes produced by \tactic.}
\label{tab:bug-study}
\scriptsize
\setlength{\tabcolsep}{4pt}
\begin{tabular}{@{}lrrr@{}}\toprule
\textbf{Backend / Component} & \textbf{\# Reported} & \textbf{\# Confirmed} & \textbf{\# Fixed} \\\midrule
CPU & 9 & 7 & 5 \\
CUDA & 7 & 5 & 5 \\
MPS & 2 & 2 & 2 \\
Other components & 13 & 12 & 8 \\\midrule
\textbf{Total} & \textbf{31} & \textbf{26} & \textbf{20} \\
\bottomrule
\end{tabular}
\end{table}

\begin{figure}[t]
    \centering
    \captionsetup[subfigure]{font=scriptsize,skip=1pt}
    \begin{subfigure}{\columnwidth}
        \centering
        \includegraphics[width=\linewidth]{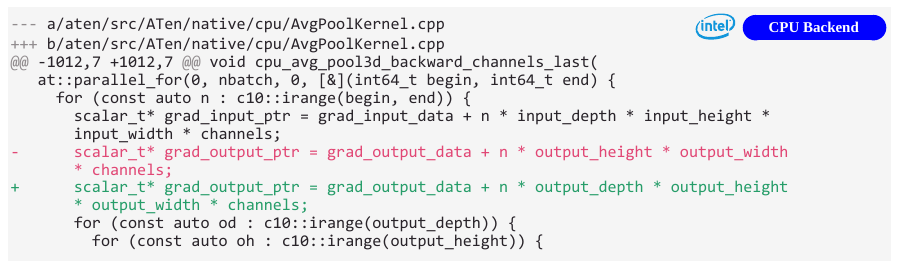}
        \caption{Patch for an incorrect batch stride in AvgPool3d backward}
        \label{fig:bug-study-cpu}
    \end{subfigure}
    \vspace{0.25em}
    \begin{subfigure}{\columnwidth}
        \centering
        \includegraphics[width=\linewidth]{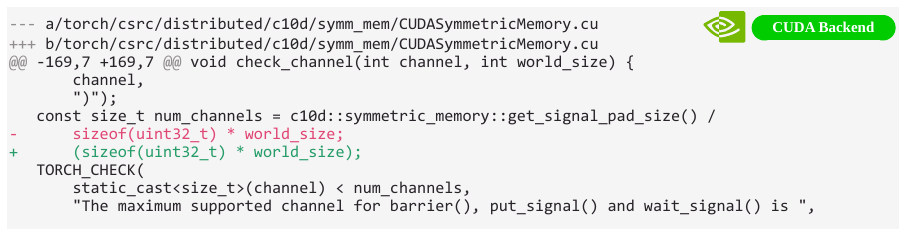}
        \caption{Patch for an operator-precedence error in channel validation}
        \label{fig:bug-study-cuda}
    \end{subfigure}
    \vspace{0.25em}
    \begin{subfigure}{\columnwidth}
        \centering
        \includegraphics[width=\linewidth]{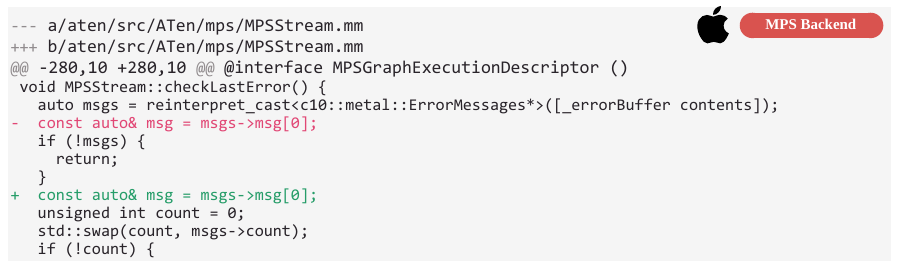}
        \caption{Patch for dereference before validating runtime state}
        \label{fig:bug-study-mps}
    \end{subfigure}
    \caption{Representative patches submitted by us after \tactic detected bugs across heterogeneous backends.}
    \label{fig:bug-study}
\end{figure}

\parabf{CPU backend.} Fig.~\ref{fig:bug-study-cpu} shows a patch submitted by us for the AvgPool3d backward kernel in the Channels Last layout. The kernel computes a pointer into \CodeIn{grad\_output\_data} for the $n$-th sample. In the Channels Last 3D layout, the batch stride should preserve the full logical shape $(D,H,W,C)$, namely \CodeIn{output\_depth * output\_height * output\_width * channels}. However, the original implementation used only \CodeIn{output\_height * output\_width * channels}, omitting the depth dimension. In SBIR, this case is represented as a shape/layout bridge from the tensor metadata of the pooling output to the backend pointer offset expression. The analyzer checks whether the address calculation preserves all dimensions in $\Phi$. Since \CodeIn{output\_depth} is dropped when constructing the target pointer, the backend violates the consistency constraint between shape and address calculation. Our patch adds the missing depth factor to preserve the tensor layout semantics. Such a bug is particularly dangerous because it may not crash immediately. Instead, it can silently read gradients from the wrong batch region and produce incorrect results.

\parabf{CUDA backend.} Fig.~\ref{fig:bug-study-cuda} presents a patch submitted by us for the distributed symmetric-memory implementation. The function \CodeIn{check\_channel} validates whether a channel index falls within the number of supported signal-pad channels. The intended capacity is computed by dividing the signal-pad size by the per-world-size storage cost (\ie \CodeIn{sizeof(uint32\_t) * world\_size}). The original code missed the parentheses and was evaluated as \CodeIn{get\_signal\_pad\_size() / sizeof(uint32\_t) * world\_size}, which can substantially overestimate the number of valid channels. SBIR captures this case as an allocation/capacity bridge from the symmetric-memory buffer layout to the CUDA-side bounds check. The violated constraint is that the maximum legal channel must be derived from the same physical buffer partitioning used by all participating processes. Our patch adds the missing parentheses so the capacity check follows the actual buffer partitioning. When this constraint is broken, the check can accept channels beyond the real per-rank capacity, creating a risk of invalid access in distributed CUDA synchronization code. This case also illustrates the complementarity of static reasoning. Dynamically triggering it requires a specific CUDA distributed-memory configuration.

\parabf{MPS backend.} Fig.~\ref{fig:bug-study-mps} shows a patch submitted by us for \CodeIn{MPSStream::checkLastError}. The original implementation creates \CodeIn{msgs} from the MPS error buffer and immediately binds \CodeIn{msgs->msg[0]} before validating whether \CodeIn{msgs} is null and before checking whether the buffer contains any messages. The patch moves the reference after the null check and count extraction. In SBIR, this bug is modeled as a control-dependency and memory-state bridge. The target entity \CodeIn{msg[0]} is safe to access only when the source runtime state satisfies \CodeIn{msgs != null} and contains at least one valid message. The original code consumes the target field before these constraints are established, violating the guard-before-dereference invariant. Compared with above two examples, this case is not a tensor shape mismatch, but it still follows the same SBIR principle. A backend implementation consumes a semantic state before preserving the required guard.

\parabf{Answer to RQ3.} \tactic detects diverse real-world bugs across heterogeneous backend implementations. These cases demonstrate that SBIR is not merely a textual summary of code snippets. It provides a common representation for checking whether backend computations preserve higher-level semantic constraints, including tensor layout, buffer capacity, and runtime memory state.

\subsection{RQ4. Ablation study}
To understand which design choices are responsible for \tactic's effectiveness, we run the four ablation variants defined in \S~\ref{sec:setup}. Table~\ref{tab:rq4-ablation} reports the number of real bugs, false alarms, and total alarms produced by each setting.
\begin{table}[htbp]
\centering
\caption{Comparison with ablation variants.}
\label{tab:rq4-ablation}
\small
\begin{tabular}{@{}lrrr@{}}
\toprule
\textbf{Technique} & \textbf{\# Bugs} & \textbf{\# False Alarms (\%)} & \textbf{\# Alarms} \\
\midrule

\tactic     & \textbf{31} & 5 (13.89\%) & 36 \\
w/o CWE     & 8           & 3 (27.27\%) & 11 \\
w/o PR      & 23          & 2 (8.00\%)  & 25 \\
w/o summary & 27          & 3 (10.00\%) & 30 \\
w/o ERG     & 17          & 26 (60.47\%) & 43 \\
\bottomrule
\end{tabular}
\end{table}

\subsubsection{Effectiveness of bug knowledge sources}
The two knowledge sources provide complementary guidance. The w/o CWE variant finds only 8 bugs, a 74.19\% reduction from \tactic, showing that general weakness knowledge helps expand the search space beyond the exact patterns already observed in \pt patches. Without CWE patterns, the variant depends more heavily on historical fixes and misses security-relevant bugs that are not close to previously collected \pt PRs. In contrast, the w/o PR variant finds 23 bugs, a 25.81\% reduction. Its false positive rate is lower than the full configuration, but this should not be read as an improvement because the variant raises fewer alarms and misses more bugs. This contrast indicates that the two heterogeneous data sources are complementary, with CWE contributing broad weakness coverage and bug-fix PRs providing framework-specific instances that make these weaknesses concrete in \pt.

\subsubsection{Effectiveness of bug summary}
The w/o summary variant finds 27 bugs, losing 4 bugs compared with the full \tactic. Its false positive rate is 10.00\%, slightly lower than the full configuration, because it also produces fewer alarms. Raw PR diffs and CWE descriptions often contain noisy implementation details or generic weakness language. The summary step normalizes them into a concise, DL-framework-specific pattern before identifier extraction and SBIR generation. Without this semantic anchor, retrieval and generation still work in many cases, but they miss some bug-relevant attributes such as shape, dtype, layout, or guard conditions. This explains why the variant remains relatively precise yet detects fewer bugs.

\subsubsection{Effectiveness of ERG}
ERG is the main guardrail against hallucinated code context during SBIR generation. Without retrieved repository context, the LLM directly synthesizes SBIRs from the bug data and summary. This variant reports 43 alarms but detects only 17 bugs, raising the false positive rate to 60.47\%, compared with 13.89\% for the full \tactic. The result shows that SBIR generation needs real code context. Otherwise, the LLM may produce plausible but invalid bridges, such as incorrect file contexts.

\parabf{Answer to RQ4.} All four components contribute to \tactic. CWE knowledge provides broad security knowledge, bug-fix PRs supply framework-specific bug patterns, summaries turn noisy source data into semantic patterns ready for analysis, and ERG is essential for keeping SBIR generation valid. 

\section{Related Work}
\label{sec:related}

\subsection{DL Framework Testing}
Testing DL frameworks for reliability and security has become an active research topic. Based on the difference in test generation, existing approaches can be categorized into traditional techniques~\cite{cradle,lemon,audee,graphfuzz,nnsmith,neuri} and techniques using LLMs~\cite{titanfuzz,fuzzgpt,whitefox,future,fuel}.

Traditional techniques adopt grammar driven methods to generate test cases for dynamic execution. CRADLE~\cite{cradle} is the first DL framework testing technique to directly run real-world models on different backends of Keras~\cite{keras}. LEMON~\cite{lemon} and AUDEE~\cite{audee} are subsequently proposed to mutate these real-world models to detect more bugs. GraphFuzz~\cite{graphfuzz} constructs computational graphs via Monte Carlo Tree Search and alleviates the tensor shape mismatch problem through padding/slicing operations. NNSmith~\cite{nnsmith} models each operator's symbolic constraint to generate valid and diverse DL models with gradient-based search. NeuRI~\cite{neuri} automates operator specification generation through inductive rule inference, enriching test diversity significantly.

In recent years, LLMs have been introduced into DL framework testing~\cite{titanfuzz,fuzzgpt,whitefox,future,fuel}. The core idea behind these techniques is that LLMs can learn implicit DL program constraints during pretraining and infer valid and diverse DL programs. \titanfuzz~\cite{titanfuzz} is a pioneering work that adopts two LLMs to generate and fill DL APIs, respectively. \fuzzgpt~\cite{fuzzgpt} generates edge test cases for fuzzing through fine tuning and in-context learning. \whitefox~\cite{whitefox} first leverages LLMs to extract optimization patterns in compilers and generate tests targeting these patterns. FUTURE~\cite{future} uses LLMs to mine bug patterns with API documents from old DL frameworks to detect similar bugs for new DL frameworks. \fuel~\cite{fuel} utilizes an analysis LLM to distill feedback information into valuable guidance, thereby helping the generation LLM produce more diverse and valid DL models.

Unlike above fuzzing approaches, \tactic is the first static technique to detect bugs in the \pt codebase, opening a new direction for DL framework testing.

\subsection{Static Analysis with LLMs}
Recent static analysis techniques using LLMs can be grouped by how they couple LLM reasoning with conventional analyzers. Early systems such as LLift~\cite{llift} use LLMs as contextual triagers for unresolved Linux kernel static analysis reports. Path-sensitive and taint-oriented systems use LLMs to supply missing specifications or reduce false positives. IRIS~\cite{iris} infers Java taint specifications and filters CodeQL data-flow paths, Artemis~\cite{artemis} combines LLM-assisted taint reasoning with inter-procedural path-sensitive analysis for server-side request forgery detection, BugLens~\cite{buglens} targets taint-style Linux kernel bug detection, and LATTE~\cite{latte} extends LLM-guided taint reasoning to static binary analysis. Other recent systems use LLMs to construct or generalize analysis facts. LLMDFA~\cite{llmdfa} decomposes data-flow analysis into source/sink identification, function summarization, and path-feasibility checking without compilation; INFERROI~\cite{InferROI} infers resource-management intentions for resource-leak detection; and Vul-RAG~\cite{vulrag} retrieves vulnerability knowledge from CVE and fix examples to guide code-level vulnerability reasoning. A further line directly synthesizes analyzer artifacts, with \knighter~\cite{knighter} generating Clang Static Analyzer checkers for Linux kernel bugs and QLCoder~\cite{qlcoder} synthesizing CodeQL queries for security vulnerabilities.

These studies show that LLMs can complement static analyzers by providing triage decisions, specifications, semantic facts, or analysis rules, but they are still designed around single-language programs, binaries, system software, or analyzer-specific artifacts. In contrast, \tactic uses SBIR to make tensor semantics explicit across Python, C++, and CUDA, enabling static reasoning with LLMs over multilingual DL framework code rather than over isolated warnings, paths, or checker/query templates.

\section{Threats to Validity}
\parabf{Internal validity.} The main internal threats come from manual decisions, baseline configuration, and LLM nondeterminism. To reduce manual bias, the first and second authors inspected the collected PRs, CWE types, and reported alarms based on their experience in bug reporting. For baseline and LLM effects, we follow the default settings in their original papers, use the same LLM backend when comparing with fuzzers that use LLMs, and process each bug or weakness item 10 times to generate multiple SBIRs before downstream analysis.

\parabf{External validity.} Our evaluation focuses on \pt, so the results may not directly generalize to all DL frameworks or software systems. However, \pt is a widely used DL framework with Python, C++, and CUDA code, heterogeneous backends, and a large real-world bug history, making it a representative target for studying cross-language tensor-semantic bugs. Our bug knowledge sources may also miss some bug types, so we combine historical \pt PRs with CWE rules to cover both framework-specific patterns and broader security-oriented weaknesses.

\section{Conclusion}
\label{sec:con}

We present \tactic, a static analysis technique using LLMs to detect cross-language bugs in tensor semantics for DL frameworks. \tactic introduces SBIR to capture how tensor attributes, guards, and runtime states propagate across Python, C++, and CUDA code, and uses LLM agents to construct and analyze these semantic bridges. Our evaluation on \pt shows that \tactic detects 31 real bugs from 36 alarms, including 30 bugs missed by DL fuzzers using LLMs. Among them, 26 bugs have been confirmed and 20 patches have been merged upstream. These results demonstrate that LLM static analysis guided by SBIR is a promising complement to existing DL framework testing techniques.



\bibliographystyle{plain}
\bibliography{refs}

\end{document}